\documentclass[11pt]{article}

\usepackage[T1]{fontenc}
\usepackage[dvipsnames,svgnames,x11names,hyperref]{xcolor}
\usepackage{amsmath}
\usepackage{amssymb, amsfonts}
\usepackage{amsfonts}
\usepackage{amsthm}
\usepackage{thmtools}

\usepackage{fullpage}

\usepackage{xspace}
\usepackage{todonotes}
\presetkeys{todonotes}{size=\footnotesize,color=green!40}{}
\numberwithin{equation}{section}
\usepackage{mdframed}
\usepackage{tikz}
\usepackage{scrextend} 
\usepackage{algorithm}
\usepackage{dsfont} 
\usepackage{prettyref}
\usepackage{algpseudocode}

\usepackage{boxedminipage}
\usepackage{bbm}

\algnewcommand\algorithmicinput{\textbf{Input:}}
\algnewcommand\INPUT{\item[\algorithmicinput]}
\algnewcommand\algorithmicoutput{\textbf{Output:}}
\algnewcommand\OUTPUT{\item[\algorithmicoutput]}
\algnewcommand\algorithmicparameters{\textbf{Parameters:}}
\algnewcommand\PARAMETERS{\item[\algorithmicparameters]}

\newcommand{\savehyperref}[2]{\texorpdfstring{\hyperref[#1]{#2}}{#2}}

\newrefformat{eq}{\savehyperref{#1}{\textup{(\ref*{#1})}}}
\newrefformat{lem}{\savehyperref{#1}{Lemma~\ref*{#1}}}
\newrefformat{def}{\savehyperref{#1}{Definition~\ref*{#1}}}
\newrefformat{thm}{\savehyperref{#1}{Theorem~\ref*{#1}}}
\newrefformat{cor}{\savehyperref{#1}{Corollary~\ref*{#1}}}
\newrefformat{cha}{\savehyperref{#1}{Chapter~\ref*{#1}}}
\newrefformat{sec}{\savehyperref{#1}{Section~\ref*{#1}}}
\newrefformat{app}{\savehyperref{#1}{Appendix~\ref*{#1}}}
\newrefformat{tab}{\savehyperref{#1}{Table~\ref*{#1}}}
\newrefformat{fig}{\savehyperref{#1}{Figure~\ref*{#1}}}
\newrefformat{hyp}{\savehyperref{#1}{Hypothesis~\ref*{#1}}}
\newrefformat{alg}{\savehyperref{#1}{Algorithm~\ref*{#1}}}
\newrefformat{sdp}{\savehyperref{#1}{SDP~\ref*{#1}}}
\newrefformat{rem}{\savehyperref{#1}{Remark~\ref*{#1}}}
\newrefformat{item}{\savehyperref{#1}{Item~\ref*{#1}}}
\newrefformat{step}{\savehyperref{#1}{step~\ref*{#1}}}
\newrefformat{conj}{\savehyperref{#1}{Conjecture~\ref*{#1}}}
\newrefformat{fact}{\savehyperref{#1}{Fact~\ref*{#1}}}
\newrefformat{fct}{\savehyperref{#1}{Fact~\ref*{#1}}}
\newrefformat{prop}{\savehyperref{#1}{Proposition~\ref*{#1}}}
\newrefformat{claim}{\savehyperref{#1}{Claim~\ref*{#1}}}
\newrefformat{relax}{\savehyperref{#1}{Relaxation~\ref*{#1}}}
\newrefformat{red}{\savehyperref{#1}{Reduction~\ref*{#1}}}
\newrefformat{part}{\savehyperref{#1}{Part~\ref*{#1}}}
\newrefformat{prob}{\savehyperref{#1}{Problem~\ref*{#1}}}
\newrefformat{ass}{\savehyperref{#1}{Assumption~\ref*{#1}}}

\newcommand{\Sref}[1]{\hyperref[#1]{\S\ref*{#1}}}

\renewcommand{\leq}{\leqslant}

\renewcommand{\geq}{\geqslant}

\newcommand{\mper}{\,.}
\newcommand{\mcom}{\,,}

\newcommand{\paren}[1]{\left(#1 \right )}

\newcommand{\set}[1]{\left\{#1\right\}}

\newcommand{\abs}[1]{\left\lvert#1\right\rvert}
\newcommand{\Abs}[1]{\left\lvert#1\right\rvert}

\newcommand{\norm}[1]{\left\lVert#1\right\rVert}

\newcommand{\defeq}{\stackrel{\textup{def}}{=}}

\newcommand{\Z}{{\mathbb Z}}

\newcommand{\R}{\mathbb R}

\newcommand{\sdp}{{\sf SDP }}

\newcommand{\opt}{{\sf OPT}}
\newcommand{\OPT}{{\sf OPT}}

\newcommand{\seteq}{\mathrel{\mathop:}=}
\newcommand{\subjectto}{\text{subject to}}

\newcommand{\Esymb}{\mathbb{E}}

\DeclareMathOperator*{\E}{\Esymb}

\newcommand{\e}{\epsilon}

\let\e\varepsilon

\newcommand{\cP}{\mathcal P}

\newcommand{\cS}{\mathcal S}

\newcommand{\cW}{\mathcal W}

\newcommand{\etal}{et~al. }
\newcommand{\bigO}{\mathcal{O}}
\newcommand{\bigo}[1]{\bigO\left(#1\right)}

\newcommand{\argmin}{{\sf argmin}}
\newcommand{\poly}{{\sf poly}}

\newcommand{\inparen}[1]{\left(#1 \right )}

\renewcommand{\epsilon}{\varepsilon}

\theoremstyle{definition}
\newtheorem{theorem}{Theorem}[section]
\newtheorem{lemma}[theorem]{Lemma}
\newtheorem{corollary}[theorem]{Corollary}

\theoremstyle{definition}
\newtheorem{definition}[theorem]{Definition}

\newtheorem{proposition}[theorem]{Proposition}
\newtheorem*{proposition*}{Proposition}
\newtheorem*{claim*}{Claim}

\newtheorem{SDP}[theorem]{SDP}
\newtheorem{Claim}[theorem]{Claim}

\newtheorem*{lemma*}{Lemma}

\newtheorem*{conjecture*}{Conjecture}

\newtheorem{fact}[theorem]{Fact}
\newtheorem*{fact*}{Fact}

\newtheorem*{hypothesis*}{Hypothesis}

\newcommand{\kvbm}{$k$-{\sf VBM}}
\newcommand{\kpart}{$k$-{\sf Part}}

\newcommand{\phiv}{\phi^{\sf V}}
\newcommand{\eps}{\varepsilon}

\newcommand{\insquare}[1]{\left[#1\right]}
\newcommand{\inbrace}[1]{\left\{ #1 \right\}}

\newcommand{\eqdef}{\seteq}

\DeclareMathOperator{\diam}{diam}

\begin{document}


\title{Planted Models for $k$-way Edge and Vertex Expansion}
\author{Anand Louis \footnote{\texttt{E-mail: anandl@iisc.ac.in}}\\
  Indian Institute of Science \\ Bangalore, India.
  \and
  Rakesh Venkat\footnote{\texttt{E-mail:rakeshvenkat@iith.ac.in}}\\
  Indian Institute of Technology, Hyderabad\\ Hyderabad, India}
\date{}

\maketitle

\begin{abstract}
Graph partitioning problems are a central topic of study in algorithms and complexity theory. 
Edge expansion and vertex expansion, two popular graph partitioning objectives,
seek a  $2$-partition of the vertex set of the graph that minimizes the considered objective.
However, for many natural applications, one might require a graph to be partitioned into $k$ parts, for some $k \geq 2$.
For a $k$-partition $S_1, \ldots, S_k$ of the vertex set of a graph $G = (V,E)$,
the {\em $k$-way edge expansion} (resp. vertex expansion) of $\set{S_1, \ldots, S_k}$ is defined as
$\max_{i \in [k]} \Phi(S_i)$,
and the balanced $k$-way edge expansion (resp. vertex expansion) of $G$ is defined as 
\[ \min_{ \set{S_1, \ldots, S_k} \in \cP_k} \max_{i \in [k]} \Phi(S_i)  \mcom  \]   
where $\cP_k$ is the set of all balanced $k$-partitions of $V$
(i.e each part of a $k$-partition in $\cP_k$ should have cardinality $\Abs{V}/k$),
and $\Phi(S)$ denotes the edge expansion (resp. vertex expansion) of $S \subset V$.
We study a natural planted model for graphs where the vertex set of a graph has a $k$-partition
$S_1, \ldots, S_k$ such that the graph induced on each $S_i$ has large expansion, but each $S_i$
has small edge expansion (resp. vertex expansion) in the graph. 	
We give bi-criteria approximation algorithms for computing the balanced $k$-way edge expansion 
(resp. vertex expansion) 
of instances in this planted model. 
\end{abstract}

\newpage

\newcommand{\kvbmparams}{\kvbm$(n,k,\eps, p, c, r, \lambda)$}
\newcommand{\kpartparams}{\kpart$(n,k,\eps, \lambda, r)$}
\newcommand{\tbigo}[1]{\tilde{O}\paren{#1}}
\newcommand{\phivksymb}{\phi^{{\sf V,k}}}
\newcommand{\phivk}[1]{\phivksymb\paren{#1}}
\newcommand{\phiksymb}{\phi^{\sf k}}
\newcommand{\phik}[1]{\phiksymb\paren{#1}}
\newcommand{\phiva}{\phi^{\sf V,a}}
\newcommand{\expansion}{\Phi}
\newcommand{\kexpansionsymb}{\expansion^{k}}
\newcommand{\kexpansion}[1]{\kexpansionsymb\paren{#1}}
\newcommand{\kparte}{$k$-Part-{\sf edge}}
\newcommand{\kpartv}{$k$-Part-{\sf vertex}}
\newcommand{\kparteparams}{\kparte$(n,k,\eps, \lambda, d, r)$}
\newcommand{\kpartvparams}{\kpartv$(n,k,\eps, \lambda, d, r)$}
\newcommand{\vol}[1]{{\sf vol}\paren{#1}}

\section{Introduction}\label{sec:intro}

The complexity of computing various graph expansion parameters are central open problems in theoretical computer science, and in spite of many decades of intensive research, they are yet to be fully understood \cite{am85,a86,lr99,arv09,fhl08,rs10}.
A central problem in the study of graph partitioning is that of computing 
the sparsest edge cut in a graph. For a graph $G=(V,E)$, we define the  
{\em edge expansion} of a set $S$ of vertices, denoted by $\phi(S)$
as 
\begin{equation}
\label{eq:defedge}
\phi(S) \defeq \frac{\Abs{E(S, V \setminus S)}}{\Abs{S} \Abs{V \setminus S}} \Abs{V} \mcom
\end{equation}
where $E(S, V \setminus S) \defeq \set{ \set{u,v} \in E | u \in S, v \in V \setminus S}$.
The edge expansion of the graph $G$ is defined as $\phi_G \defeq \min_{S \subset V} \phi(S)$. 
Related to this is the notion of the vertex expansion of a graph.
For a graph $G=(V,E)$, we define the  
{\em vertex expansion} of a set $S$ of vertices, denoted by $\phiv(S)$
as
\begin{equation}
\label{eq:defvertex}
\phiv(S) \defeq \frac{\Abs{N(S) \cup N(V \setminus S)}}{\Abs{S} \Abs{V \setminus S}} \Abs{V} \mcom 
\end{equation}
where
$N(S) \defeq \set{v \in V \setminus S | \exists u \in S \textrm{ such that} \set{u,v} \in E }$.
The vertex expansion of the graph $G$ is defined as $\phiv_G \defeq \min_{S \subset V}\phiv(S)$. 
A few other related notions of vertex expansion have been studied in the literature, we 
discuss them in \prettyref{sec:related}.
We also give a brief description of related works in \prettyref{sec:related}.

\paragraph*{Graph $k$-partitioning.}
The vertex expansion and edge expansion objectives seek a  $2$-partition of the vertex set of the graph.
However, for many natural applications, one might require a graph to be partitioned into $k$ parts, for some $k \geq 2$.
Let us use $\expansion$ to denote either $\phi$ (edge expansion) or $\phiv$ (vertex expansion).
For a $k$-partition $S_1, \ldots, S_k$ of the vertex set,
the {\em $k$-way edge/vertex expansion} of $\set{S_1, \ldots, S_k}$ is defined as
\[ \kexpansion{S_1, \ldots, S_k} \defeq \max_{i \in [k]} \expansion(S_i) \mcom \]
and the {\em $k$-way edge/vertex expansion} of $G$ is defined as 
\[ \kexpansionsymb_G \defeq \min_{ \set{S_1, \ldots, S_k} \in \cP_k} \kexpansion{S_1, \ldots, S_k} \mcom  \]   
where $\cP_k$ is the set of all $k$-partitions of the vertex set.
Optimizing these objective function is useful when one seeks a $k$-partition
where each part has small expansion. 
The edge expansion version of this objective has been studied in 
\cite{lrtv12,lm14a,lot14}, etc., and the vertex expansion version of this 
objective has been studied in \cite{cltz18}; 
see 
\prettyref{sec:related} for a brief summary of the related work.

For many NP-hard optimization problems, simple heuristics work very well in practice,
for e.g. SAT \cite{bp99}, sparsest cut \cite{kk95,kk98}, etc. 
One possible explanation for this phenomenon could be that instances arising in practice
have some inherent structure that makes them ``easy''. 
Studying natural random/semi-random families of instances, and instances with planted solutions has
been a fruitful approach towards understanding the structure of easy 
instances, and in modelling instances arising in practice,  
especially for graph partitioning problems \cite{m01,mmv12,mmv14,LV18}
(see \prettyref{sec:related} for a brief survey).
Moreover, studying semi-random and planted instances of a problem can be used to better
understand what aspects of a problem make it ``hard''.
Therefore, in an effort to better understand the complexity of graph $k$-partitioning problems,
we study the $k$-way edge and vertex expansion of a natural planted model of instances. 
We give bi-criteria approximation algorithms for instances from these models.

\subsection{$k$-way planted models for expansion problems} 
\label{sec:model}
We study the following model of instances. 

\begin{definition}[\kparte]
\label{def:kparte}
An instance of \kparteparams~is generated as follows.
\begin{enumerate}
\item 
\label{step:def-e-expander}
Let $V$ be a set of $n$ vertices. Partition $V$ into $k$ sets $\set{S_1, S_2, \ldots S_k}$, with 
$\abs{S_t} = n/k$ for every $t \in [k]$. For each $t \in [k]$, 
add edges between arbitrarily chosen pairs of vertices in $S_t$ to form an arbitrary roughly $d$-regular 
(formally, the degree of each vertex should lie in $[d, rd]$) 
graph of spectral gap (defined as the second smallest eigenvalue of the normalized
Laplacian matrix of the graph, see \prettyref{sec:notation} for definition) at least $\lambda$. 

\item 
\label{step:def-e-sparsecut}
For all $i,j \in [k]$, add edges between arbitrarily chosen pairs of vertices in $S_i \times S_j$
such that $\phi_G(S_i) \leq \e r d \ \forall i \in [k]$. 

\item (Monotone Adversary)
For each $t \in [k]$, add edges between any number of arbitrarily chosen pairs of vertices within $S_t$. 

\end{enumerate}
Output the resulting graph $G$.
\end{definition}

Analogously, we define the vertex expansion model. 

\begin{definition}[\kpartv]
\label{def:kpartv}
An instance of \kpartvparams~is generated as follows.
\begin{enumerate}
\item 
\label{step:def-v-expander}
Let $V$ be a set of $n$ vertices. Partition $V$ into $k$ sets $\set{S_1, S_2, \ldots S_k}$, with 
$\abs{S_t} = n/k$ for every $t \in [k]$. For each $t \in [k]$, 
add edges between arbitrarily chosen pairs of vertices in $S_t$ to form an arbitrary roughly $d$-regular 
(formally, the degree of each vertex should lie in $[d, rd]$) 
graph of spectral gap (defined as the second smallest eigenvalue of the normalized
Laplacian matrix of the graph, see \prettyref{sec:notation} for definition) at least $\lambda$. 

\item 
\label{step:def-v-sparsecut}
For each $t \in [k]$, partition $S_t$ into $T_t$ and $S_t \setminus T_t$ such that
$\Abs{T_t} \leq \e n /k$. 
Add edges between any number of arbitrarily chosen pairs of vertices in $\cup_{i \in [k]}T_i$.

\item (Monotone Adversary)
For each $t \in [k]$, add edges between any number of arbitrarily chosen pairs of vertices within $S_t$.

\end{enumerate}
Output the resulting graph $G$.
\end{definition}

The only difference between \kparte~and \kpartv~is in the expansion of the sets. 
In \prettyref{step:def-e-sparsecut} of \prettyref{def:kparte}, we ensured that
$\phi(S_i) \leq \e r d\ \forall i \in [k]$\footnote{Since 
$\phi(S)$ measures the weight of edges leaving $S$
(see \prettyref{eq:defedge}), it is often more useful to compare edge expansion to 
some quantity related to the degrees of the vertices inside $S$. Therefore, 
in \prettyref{step:def-e-sparsecut} of \prettyref{def:kparte}, we require
$\phi(S_i) \leq \e r d\ \forall i \in [k]$, instead of 
$\phi(S_i) \leq \e\ \forall i \in [k]$.}.
In \prettyref{step:def-v-sparsecut} of \prettyref{def:kpartv}, the definition 
ensures that $\phiv(S_i) \leq \e k \ \forall i \in [k]$.

Both these models can be viewed as the generalization to $k$-partitioning 
of models studied in the literature for $2$-partitioning problems 
for edge expansion \cite{mmv12}, etc. and vertex expansion \cite{LV18}, etc.
These kinds of models can be used to model communities in networks, where $k$ is
the number of communities.
The intra-community connections are typically stronger than the inter-community
connections. 
This can be modelled by requiring
$S_i$ to have large expansion (see \prettyref{thm:kparte} and \prettyref{thm:kpartv}
for how large a $\lambda$ is needed compared to $\e$).
Our work for $k>2$ can be used to study more general models of communities than
the case of $k=2$.

\subsection{Our Results} \label{sec:results}

We give bi-criteria approximation algorithms for the instances generated from
the \kparte~and \kpartv~models. We define $\opt$ as follows
\begin{align*}
 \opt \defeq \min_{ \set{P_1, \ldots, P_k} \in \widetilde{\cP}_k} \kexpansion{P_1, \ldots, P_k} \mcom 
\end{align*}
where $\expansion$ is $\phi$ for \kparte, and $\phiv$ for \kpartv, and
$\widetilde{\cP}_k$ is the set of all \emph{balanced} $k$-partitions of the vertex-set, 
i.e. for each $\set{P_1, \ldots, P_k} \in \widetilde{\cP}_k$, we have $\Abs{P_i} = n/k\ \forall i \in [k]$.
We note that in \kparte, $\opt \leq \eps rd$, and in \kpartv, $\opt \leq \eps k$.

\begin{theorem}
\label{thm:kparte}
There exist universal constants $c_1, c_2 \in \R^+$ satisfying the following:
there exists a polynomial-time algorithm that takes as input a graph from the class 
\kparteparams~with $\eps \leq  \lambda/(800 k r^3)$, and outputs $k$ disjoint sets of 
vertices $W_1, \ldots ,W_k \subseteq V$,  that for each $i\in[k]$ satisfy:
\begin{enumerate}
\item $\abs{W_i} \geq c_1 n/k$, 
\item  $\phi(W_i) \leq  c_2 k \opt$.
\end{enumerate}
\end{theorem}

\begin{theorem}
\label{thm:kpartv}
There exist universal constants $c_1, c_2 \in \R^+$ satisfying the following:
there exists a polynomial-time algorithm that takes as input a graph from the class 
\kpartvparams~with $\eps \leq  \lambda/(800 k r^3)$, and outputs $k$ disjoint sets of 
vertices $W_1, \ldots ,W_k \subseteq V$,  that for each $i\in[k]$ satisfy:
\begin{enumerate}
\item $\abs{W_i} \geq c_1 n/k$, 
\item  $\phiv(W_i) \leq c_2 k\opt$.
\end{enumerate}
\end{theorem}

Note when $k = \bigo{1}$, \prettyref{thm:kparte} and \prettyref{thm:kpartv} guarantee 
constant factor bi-criteria approximation algorithms.
The currently best known approximation guarantees for general instances (i.e. worst case approximation guarantees)
of $k$-way edge expansion problems are of the form $\bigo{\OPT \sqrt{\log n} f_1(k)}$ or $\bigo{\sqrt{\OPT} f_2(k)}$
where $f_1(k), f_2(k)$ are some functions of $k$, and
 the currently best known approximation guarantees for general instances (i.e. worst case approximation guarantees)
of $k$-way vertex expansion problems are of the form $\bigo{\OPT \sqrt{\log n} f_3(k)}$ or $\bigo{\sqrt{\OPT} f_4(k,d)}$
where $f_3(k)$ is some functions of $k$ and $f_4$ is some function of $k$ and the maximum vertex degree $d$.
We survey these results in \prettyref{sec:related}.
Note that our bi-criteria approximation guarantees in \prettyref{thm:kparte} and \prettyref{thm:kpartv} 
are multiplicative approximation guarantees and are independent of $n$.

The above theorem shows that it is possible to produce $k$ disjoint subsets, each of size $\Omega(n/k)$, 
each with expansion a factor $k$ away from that of the planted partition. While this may not form a partition of the vertex set, 
it is not difficult to show that with a loss of a factor of $k$, we can indeed get a true partition. 
This idea of moving from disjoint sets to a partition is well-known, and has been used before in other works (for e.g., \cite{lot14}). 

\begin{corollary}
\label{cor:approx-recovery-1}
There exist universal constants $c_1, c_2 \in \R^+$ satisfying the following:
there exists a polynomial-time algorithm that takes as input a graph from 
\kparteparams~(resp. \kpartvparams) with $\eps \leq  \lambda/800 k cr^3$, and outputs a $k$-partition 
$\mathcal{P} = \set{P_1, \ldots, P_k}$  of $V$ such that:
\begin{enumerate}
\item For each $i \in [k]$, $\abs{P_i} \geq c_1 n/k$,
\item For each $i \in [k]$, $\phi(P_i) \leq  c_2 k^2 \opt$ 
\hspace{0.1cm} (resp. $\phiv(P_i) \leq c_2 k^2 \opt $). 
\end{enumerate} 
\end{corollary}

We note that the above result approximates the $k$-way expansion of the best \emph{balanced} partition in $G$. 
The proofs of the above results are given in \prettyref{sec:approximate-recovery-proof}.

\subsection{Proof Overview}

For proving \prettyref{thm:kparte} and \prettyref{thm:kpartv} we use an SDP relaxation 
(see \prettyref{sec:sdp}) similar to the one used by \cite{lm14a,mmv16}, etc.
For the case when $k=2$, \cite{mmv12,LV18} used slightly different SDP constraints, and showed that 
when $S_1$ and $S_2$ contain large edge expanders, the set of SDP solution vectors 
$\set{u_i : i \in V}$ contain two sets $L_1,L_2$ such that 
$\Abs{L_1}, \Abs{L_2} = \Omega(n)$, $L_1$ and $L_2$ have small diameter, and the distance between 
$L_1$ and $L_2$ is $\Omega(1)$.  The core of our analysis can be viewed as proving an analogue of this for $k>2$ 
(\prettyref{prop:sdp-is-clustered-2}), however, this requires some new ideas. 
For $i \in [k]$, let $\mu_i$ denote the mean of the vectors corresponding to the vertices in $S_i$.
We use the expansion within $S_i$'s together with the SDP constraints to show that 
for $i,j \in [k]$, $i \neq j$, each $\mu_i$ must have $\Omega(n/k)$ vertices sufficiently close to it, 
and that $\mu_i$ and $\mu_j$ must be sufficiently far apart.
This can be used to show the existance of $k$ such sets $L_1, \ldots, L_k$, such that for each $i \in [k]$, $L_i$ 
has sufficiently small diameter and $L_i$ is  sufficiently
far from $L_j$ $\forall j \neq i$.
The proof of our structure theorem is similar in spirit to the proof of structure theorem of \cite{psz17},
but our final guarantees are very different, we discuss their work in more detail in \prettyref{sec:related}.

If we can compute $k$ such sets $L_1, \ldots, L_k$, 
then using standard techniques, we can recover $k$ sets having small expansion.
In the case of $k=2$, one could just guess a vertex from each these sets,
and compute the two sets satisfying our requirements using standard techniques. 
For $k>2$, 
guessing a vertex from each of the balls around $\mu_i$ would also suffice to compute sets $L_1, \ldots, L_k$
satisfying our requirements. However, doing this naively would take time $O(n^k)$.
To obtain an algorithm for this task whose running time is $\bigo{{\sf poly}(n,k)}$,
we use a simple greedy algorithm (Algorithm 1) to iteratively compute the sets $L_i$ such that
$L_i$ has sufficiently small diameter and is sufficiently far from $L_j$ for all $j<i$. 
To ensure that this approach works, one has to ensure that at the start of iteration $i+1$, 
the set of SDP vectors for the vertices in 
$V \setminus \cup_{j = i}^{i} L_i$ has at least $k-i$ clusters each of size $\Omega(n/k)$ and having small 
diameter. 
We use our structural result to prove that this invariant holds 
in all iterations of the algorithm. 

\subsection{Related Work}
\label{sec:related}

\cite{LV18} studied the $2$-way vertex-expansion in \kpartv~for $k=2$, and gave a constant factor bi-criteria approximation algorithm. Our proofs
and results can be viewed as generalizing their result to $k>2$. They also studied a stronger semi-random model, and gave 
an algorithm for exact recovery (i.e. a $1$-approximation algorithm) w.h.p.
\cite{mmv12} studied the $2$-way edge-expansion in a model similar to \kparte~for $k=2$, and gave a constant 
factor bi-criteria approximation algorithm. Our proofs
and results can be viewed as generalizing their result to $k>2$. 
\medskip

\noindent \textbf{$k$-partitioning problems.} \\
The minimum $k$-cut problem asks to find a $k$-partition of the vertex set
which cuts the least number of edges; \cite{sv95,nr01,rs08} all gave $2$-approximation algorithms
for this problem. A number of works have investigated $k$-way partitioning in the context of edge expansion.
Bansal \etal \cite{bfk11} studied the problem of computing a $k$-partitioning 
$S_1, \ldots, S_k$ of the vertex set  such that $\Abs{S_i} = n/k$ for each $i \in [k]$, which minimizes 
$\max_{i \in [k]} \Abs{E(S_i, V \setminus S_i)}$. They give an algorithm
which outputs a $k$-partition of the vertex set $T_1, \ldots, T_k$ such that
$\Abs{T_i} \leq (2+\e) n/k$, and $\max_{i \in [k]} \Abs{E(T_i, V \setminus T_i)} \leq 
\bigo{\sqrt{\log n \log k}} \opt$, where $\opt$ denotes the cost of the optimal solution.
There are also many connections between graph partitioning problems and graph eigenvalues.
Let $ 0 = \lambda_1 \leq \lambda_2 \leq \ldots \leq \lambda_n$ denote the eigenvalues of 
the normalized Laplacian matrix of the graph.
Typically, a different but related notion of edge expansion is used, which is defined as follows.
\[ \phi'(S) \defeq \frac{\Abs{E(S, V \setminus S)}}{\min \set{ {\sf vol}(S), {\sf vol}\paren{V \setminus S}}},    \]
where ${\sf vol}(S)$ is defined as the sum of the degrees of the vertices in S.
\cite{lrtv11} gave an algorithm to find a $k$-partition
which cuts at most $\bigo{\sqrt{\lambda_k} \log k}$ fraction of the edges. 
\cite{lot14,lrtv12} showed that for any $k$ non-empty disjoint subsets $S_1, \ldots, S_k \subset V$,
$\max_{i \in [k]} \phi'(S_i) = \Omega(\lambda_k)$.
\cite{lot14} (see also \cite{lrtv12,lm14a}) gave an algorithm to find a $(1-\e)k$
partition $S_1, \ldots, S_{(1 - \e)k}$ of the vertex set satisfying 
$\max_{i} \phi'(S_i) = \bigo{ \paren{1/\e^3} \sqrt{\lambda_k \log k}}$
for any $\e>0$,  and a collection of 
$k$ non-empty, disjoint subsets $S_1, \ldots, S_k \subset V$ satisfying 
$\max_{i} \phi'(S_i) = \bigo{k^2\sqrt{\lambda_k}}$. \cite{lm14a} gave an algorithm to find a partition of $V$ into $(1-\eps)k$ disjoint subsets $S_1, S_2, \ldots, S_{(1-\eps)k}$, such that $\phi'(S_i) \leq \bigo{\sqrt{\log n \log k} \opt}$.

Given a parameter $\delta$, the small-set edge expansion problem asks to compute the set $S \subset V$ have 
the least edge expansion among all sets of 
cardinality at most $\delta \Abs{V}$ (or volume at most $\delta \vol{V}$).
Bansal \etal \cite{bfk11} and Raghavendra \etal \cite{rst10}
gave a bi-criteria approximation algorithm for the small-set edge expansion problem.
\cite{lm14a} gave an algorithm that outputs $(1- \e)k$ partition
$S_1, \ldots, S_{(1 - \e)k}$ such that $\max_i \phi'(S_i) = \bigo{{\sf poly}(1/\e) \sqrt{\log n \log k}~\opt}$,
where $\opt$ is least value of $\max_{i \in [k]} \phi'(S_i)$ over all $k$-partitions 
$S_1, \ldots, S_k$ of the vertex set. 
\cite{lm14a} also studied a balanced version of this problem, and gave bi-criteria approximation
algorithms.

Let $\rho_k(G)$ denote $\min_{S_1, \ldots, S_k} \max_{i \in [k]} \phi'(S_i)$ 
where the minimum is over sets of $k$ non-empty disjoint subsets $S_1, \ldots, S_k \subset V$.
Kwok \etal \cite{kll13} showed that
for any $l>k$, $\rho_k(G) = \bigo{l k^6 \lambda_k/\sqrt{\lambda_l}}$.
They also gave a polynomial time algorithm to compute non-empty disjoint sets $S_1, \ldots, S_k \subset V$
satisfying this bound.
Combining this with the results of \cite{lot14,lrtv12},
we get a $\bigo{l k^6/\sqrt{\lambda_l}}$ approximation to
the problem of computing $k$ non-empty disjoint subsets $S_1, \ldots, S_k \subset V$
which have the least value of $\max_{i \in [k]} \phi'(S_i)$. 
Here the approximation factor depends on $\lambda_l$, but even in the best case when $\lambda_{l} = \Omega(1)$
for some $l = O(k)$, the expression for the approximation guarantee reduces to $\bigo{k^7}$.
They also show that for any $l>k$ and any $\e > 0$, there is a polynomial time algorithm to compute non-empty disjoint subsets
$S_1, \ldots, S_{(1-\e)k} \subset V$ such that 
$\max_{i \in [(1-\e)k]} \phi'(S_i) = \bigo{\paren{\paren{l \log^2 k}/\paren{{\sf poly}(\e) k}} \lambda_k/\sqrt{\lambda_l}}$.

Peng \etal \cite{psz17} define the family of well clustered graphs to be those graphs for which 
$\lambda_{k+1}/\rho_k(G) = \Omega(k^2)$
(their structure theorem requires this ratio to be $\Omega(k^2)$, their algorithms require the 
separation to be larger, i.e. $\Omega(k^3)$) . 
They show that for such graphs, using the bottom $k$ eigenvectors of the normalized Laplacian matrix,
one can compute a $k$-partition which is close to the optimal $k$-partition for $k$-way edge expansion.
They measure the closeness of their solution to the optimal solution in terms of the volume of the 
symmetric difference between the solution returned by their algorithm and the optimal solution.
They start by showing that the vertex embedding of the graph into the $k$-dimensional space consisting 
of the bottom-$k$ eigenvectors is clustered.
Our technique to prove our main structural result \prettyref{prop:sdp-is-clustered-2}, which shows that the 
SDP solution is clustered, is similar in spirit.
Firstly, we note that the results of \cite{psz17} apply to edge expansion problems and not vertex expansion
problems. 
Moreover, due to the action of the monotone adversary, the $\lambda_{k+1}$ of instances from \kparte~
could be very small in which case the results of \cite{psz17} wouldn't be applicable.  

\cite{cltz18} showed that for a hypergraph $H = (V,E)$, there exist 
$(1-\e)k$ disjoint subsets $S_1, \ldots, S_{(1-\e)k}$ of the vertex set such that 
$\max_i \phi(S_i) = \bigo{k^2 \poly\log(k)/e^{1.5}} \sqrt{\gamma_k \log r}$, where $r$ is the 
size of the largest hyperedge, $\phi(S)$ denotes the hypergraph expansion of a set 
of vertices $S$, $\gamma_k$ is the $k$th smallest eigenvalue of the 
hypergraph Laplacian operator (we refer the reader to \cite{cltz18} for the definition
of $\phi(\cdot)$, $\gamma_k$, etc.)
Combining these ideas from \cite{cltz18} with the ideas from \cite{lm16}, we believe
it should be possible to obtain an algorithm that outputs $(1- \e)k$ disjoint subsets
$S_1, \ldots, S_{(1 - \e)k}$ such that $\max_i \phi(S_i) = \bigo{k^2 \poly\log(k){\sf poly}(1/\e)} \sqrt{\log n}\, \opt$,
where is $\opt$ is least value of $\max_{i \in [k]} \phi(S_i)$ over all $k$-partitions 
$S_1, \ldots, S_k$ of the vertex set. 
Using a standard reduction from vertex expansion in graphs to hypergraph expansion,
we get analogs of the above mentioned results for vertex expansion in graphs.

\medskip

\noindent \textbf{Vertex Expansion.}
An alternative, common definition of vertex expansion that has been studied in the literature is  
$\phiva(S) \defeq  \paren{ \Abs{V} \Abs{N(S)}/\paren{ \Abs{S} \Abs{V \setminus S}}}$, 
and as before, $\phiva_G \defeq \min_{S \subset V} \phiva(S)$.
As Louis \etal \cite{lrv13} show,  the computation $\phiv_G$ and $\phiva_G$ is equivalent upto constant factors.

Feige \etal \cite{fhl08} gave a $\bigo{\sqrt{\log n}}$-approximation algorithm for  computing the vertex expansion of a graph. 
Bobkov \etal~\cite{bht00} gave a Cheeger-type inequality for vertex expansion in terms of a parameter $\lambda_\infty$, which plays a role similar to $\lambda_2$ in edge-expansion. 
Building on this, Louis \etal \cite{lrv13} gave an SDP based algorithm to compute a set having vertex expansion at most $\bigo{\sqrt{\phiv_G \log d}}$ in graphs having vertex degrees at most $d$. 
This bound is tight upto constant factors \cite{lrv13} assuming the SSE hypothesis. Louis and Makarychev~\cite{lm16} gave a bi-criteria approximation for small-set vertex expansion.

\noindent \textbf{Edge Expansion.}
Arora \etal \cite{arv09} gave a $\bigo{\sqrt{\log n}}$-approximation algorithm for computing the edge expansion of a graph.
Cheeger's inequlity \cite{am85,a86} says that $\lambda_2/2 \leq \min_{S \subset V} \phi'(S) \leq \sqrt{2 \lambda_2}$.

\noindent \textbf{Stochastic Block Models and Semi-Random Models.}
Stochastic Block Models (SBMs) are randomized instance-generation models based on the edge expansion objective and have been intensively studied in various works, starting with \cite{hll83, b87,js98}.  The goal is to identify and recover communities in a given random graph, where edges within communities appear with a probability $p$ that is higher than the probability $q$ of edges across communities. Both exact and approximate recovery guarantees for SBMs have been investigated using various algorithms~\cite{m01,mns14a,m14,abh16,mns15a,mns15b}, leading to the resolution of a certain conjecture regarding for what range of model parameters are recovery guarantees  are possible. While the above results deal mostly with the case of SBMs with two communities, $k$-way SBMs (for $k>2$ communities) have been studied in recent works \cite{as15a,as15b,abkk15}.    

Semi-Random Models allow instance generation using a combination of both random edges and some amount of monotone adversarial action (i.e. not change the underlying planted solution). SDP-based methods seem to work well in this regard, since they are robust to such adversarial action. 
Many variants of semi-random models for edge expansion have been studied in literature. Examples include works due to Feige and Kilian \cite{fk01}, Guedon and Vershynin \cite{gv16}, Moitra \etal\cite{mpw16}, and  Makarychev \etal \cite{mmv12, mmv14, mmv16}. \cite{mmv16} also allows for a small amount of non-monotone errors in their model. These works give approximate and exact recovery guarantees for a range of parameters in their respective models.

\section{Preliminaries and Notation}

\subsection{Notation} \label{sec:notation}
We denote graphs by $G=(V,E)$, where the vertex set $V$ is identified with $[n]\defeq \inbrace{1, 2, \ldots n}$. The vertices are indexed by $i,j$. 
 For any $S \subseteq V$, we denote the induced subgraph on $S$ by $G[S]$. Given $i \in V$ and $T \subseteq V$, define $N_T(i) \defeq \set{j \in T ~:~ \set{i,j} \in E}$, and $N(i) = N_V(i)$. 
 Given the normalized Laplacian $\mathcal{L} = I- D^{-1/2} A D^{-1/2}$, the \emph{spectral gap} of $G$ denoted by  $\lambda$, is the second-smallest eigenvalue of $\mathcal{L}$. \emph{Spectral expanders} are a family of graphs with $\lambda$ at least some constant (independent of the number of vertices in $G$).

Specific to graphs $G$ generated in the \kpartv~and \kparte~models, let $\cS = \inbrace{S_1, \ldots, S_t}$ be the collection of sets for any $i\in V$, let  $S(i)$ denote the set $S \in \cS$ such that $i\in S$.  For a single subset $W\subseteq V$, we define $\partial{W}=\set{i\in W\,:\, \exists j \notin W \text{ with } j \in N(i)  } \cup \set{i\notin W\,:\, \exists j \in W \text{ with } j \in N(i)  } $, i.e., the \emph{symmetric vertex} boundary of the cut $(W, V\setminus W)$. We let $E(\partial S)$ be the edges going across the cut $(S, V\setminus S)$, for any $S \subseteq V$. Given any $k$-partition of the vertex set $\cW = \set{W_1, \ldots, W_k}$, we define $\partial \cW = \cup_{i\in [k]} \partial{W_i}$ to be the set of boundary vertices on this partition, and $E(\partial \cW) = \cup_{i\in [k]} E(\partial{W_i}) $ to be the edges across this partition.

\subsection{SDP for $k$-way edge and vertex expansion} \label{sec:sdp}

Our algorithms for both \kparte~and \kpartv~models use a natural semi-definite programming (SDP) relaxation for $k$-way expansion.  The objective function we use is the `min-sum' objective in each case. For  \kpartv~, it looks to minimize the number of boundary vertices in a balanced $k$-way partition of the vertex set, and correspondingly in \kparte ,  the total number of edges across a balanced $k$-way partition of the vertex set.

\smallskip

For the \kparte~model, we use the following SDP relaxation.

\medskip

\begin{boxedminipage}{1.0\linewidth}
\begin{SDP}[Primal] \kparte
\label{sdp:primal-edge}
\[\min_{U} \qquad \frac{1}{2}\sum_{i,j\in E} U_{ii} + U_{jj} -2 U_{ij} \]\\
\subjectto
\begin{align*}
& \qquad  \qquad  U_{ii} = 1 & \forall i \in V \\
& \qquad \qquad  U_{ij} \geq 0  & \forall i,j \in V \\
& \qquad \qquad \sum_j U_{ij} = n/k  &\forall i \in V\\
& \qquad \qquad U_{jj} \geq  U_{ij}+ U_{jk} - U_{ik} & \forall i,j,k \in V\\
&  \qquad \qquad U \succeq 0
\end{align*}
\end{SDP}
\end{boxedminipage}%

\medskip

\begin{boxedminipage}{1.0\linewidth}
\begin{SDP}[Primal] \kpartv
\label{sdp:primal}
\[\min_{U} \qquad \sum_{i\in V} \eta_i\]
\subjectto
\begin{align*}
& \quad   \qquad \eta_i \geq  U_{ii}+U_{jj} - 2 U_{ij}  &\forall i, \forall j\in N(i)\\
& \qquad  \qquad  U_{ii} = 1 & \forall i \in V \\
& \qquad \qquad  U_{ij} \geq 0  & \forall i,j \in V \\
& \qquad \qquad \sum_j U_{ij} = n/k  &\forall i \in V\\
& \qquad \qquad U_{jj} \geq  U_{ij}+ U_{jk} - U_{ik} & \forall i,j,k \in V\\ 
&  \qquad \qquad U \succeq 0
\end{align*}
\end{SDP}
\end{boxedminipage}%
\medskip

The intended integral solution for $U$ in the SDP relaxation (\prettyref{sdp:primal}, \prettyref{sdp:primal-edge}) for either model is $U_{ij}= 1$, if $i,j$ lie in the same subset in the optimal $k$-partition of $V$, and $0$ otherwise.  We can alternatively view the SDP variables as a set of vectors $\set{u_i \in \R^n}_{i \in V}$, satisfying $u_i^T u_j = U_{ij}$. These can be obtained by the Cholesky decomposition of the matrix $U$.  Notice that the constraint $\sum_{j} U_{ij} = n/k$ in the relaxations above is specific to $k$-way partitions  with exactly $n/k$ vertices in each partition, and hence is satisfied by both models for the integral solution.  The second-to-last set of constraints in either SDP are called $\ell_2^2$ triangle inequalities, and can be rephrased in the language of vectors as:
\begin{align} 
\norm{u_i - u_j}^2 + \norm{u_k- u_j}^2 \geq  \norm{u_i-u_k}^2 \qquad \forall i,j,k \in V \label{eq:triang-ineq}
\end{align}
It is easy to verify that these are satisfied by the ideal integral solution, corresponding to $u_i = e_{t}$, where $i \in S_t$. 

For \kparte, for every edge across the partition we accumulate a value of $1$ in the SDP objective in the integral solution. Since every $S_t$ has $\phi(S_t) \leq \eps rd$, we have:
\begin{align*}
\abs{E(\partial S_t)} &\leq~ \eps rd \frac{n}{k} \cdot (1-\frac{1}{k})~ \leq \eps r d \frac{n}{k}\\
\implies  2 \abs{\cup_{t=1}^k E(\partial S_t)} &\leq \eps r d n
\end{align*}
 Since the number of edges going across the partition is at most\footnote{we use a slightly loose upper bound for convenience, to match up parameters in our proofs with the \kpartv~model} $\eps rd n$, this is an upper bound on the optimum of \prettyref{sdp:primal-edge}. 
 
 For \kpartv~, the integral solution will further set, $\eta_i = 2$ for any boundary vertex $i$ of the partition $\cS$, and $\eta_i = 0$ if $i$ is not a boundary vertex, yielding a primal objective value of $2 \eps n$. Thus, the optimal value of \prettyref{sdp:primal} is at most $2\eps n$. 

Furthermore, if $\opt$ is as defined in \prettyref{sec:results}, then in either case we have that $\sdp\leq \opt\cdot n$.

 We introduce some notation regarding the SDP solution vectors $\set{u_i}_{i \in V}$ that will be useful for proofs. Let $d(i,j) \defeq \norm{u_i - u_j}^2$. Due to inequalities \prettyref{eq:triang-ineq}, $d(\cdot, \cdot)$ is a metric. Given a set $L \subseteq V$, define $d(i,L) \defeq \min_{j \in L} d(i,j)$. The $\ell_2^2$ diameter of $L$ is $\diam(L)=\max_{i,j \in L} d(i,j)$. A ball of $\ell_2^2$ radius $a$ around a point $x \in \R^n$ is defined as $B(x,a) \defeq \set{j \in V: d(j, x) \leq a}$.

\medskip
Further proof-specific notations are defined as and when they are needed in the respective sections. 
\section{Bi-criteria Guarantees in the Planted Model}\label{sec:approximate-recovery-proof}
We now give a proof of \prettyref{thm:kparte}, ~\prettyref{thm:kpartv} and \prettyref{cor:approx-recovery-1}. The main idea is to show that the SDP solution is clustered around $k$ disjoint balls, each of which have a significant overlap with a distinct $S_i$, for $i \in [k]$. We can then extract out $k$ sets greedily using an $\ell_1$ line embedding.

\smallskip
In what follows, it is convenient to view the variables in the primal SDP as being vectors $u_i \in \R^n$ for each $i \in V$ that satisfy $u_i^T u_j = U_{ij}$.

\smallskip


\subsection{Preliminary Lemmas}

\begin{lemma}\label{lem:cluster-suffices}
Let $\delta \leq 1/100$ and $\alpha \leq 1$ be real numbers. Let $\{u_i\}_{i \in V}$ be a feasible SDP solution vector set for \prettyref{sdp:primal-edge} or \prettyref{sdp:primal}.   
Suppose there exists a set $L\subseteq V$ that satisfies: 
\begin{enumerate}
\item[(a)] $\abs{L} \geq \alpha n$ 
\item[(b)] $\diam(L) \leq \delta$.
\end{enumerate}
We have:
\begin{enumerate}
\item[(a)] (Edge) If $\{u_i\}_{i\in V}$ is an optimal solution to \prettyref{sdp:primal-edge} with objective value $\beta n$, then there exists an $i \in L$, and $a \in [\delta, 1/50]$ such that  $W\defeq B(i,a)$ satisfies $\phi(W) \leq \bigO(\beta /\alpha)$.
  
 \item[(b)] (Vertex) If $\{u_i\}_{i\in V}$ is an optimal solution to \prettyref{sdp:primal} with objective value $\beta n$, then there exists an $i \in L$, and $a \in [\delta, 1/50]$ such that  $W\defeq B(i,a)$ satisfies $\phiv(W) \leq \bigO(\beta /\alpha)$.
  \end{enumerate}
\end{lemma}

 Part (a) of the above lemma follows from standard arguments in edge-expansion literature.  Part (b) is a slight modification of \cite[Lemma 3.1]{LV18} \footnote{References to the results and proofs in \cite{LV18} are with respect to the full version of that paper, available currently as an arXiv preprint.}. We defer both proofs to Appendix~\ref{sec:appendix-approx-recovery}.

We next show that if  the SDP solution is clustered into $k$ disjoint, well-separated balls of small diameter, then we can iteratively use \prettyref{lem:cluster-suffices} to find $k$ disjoint sets, each with small vertex or edge expansion.

\begin{lemma} \label{lem:k-clusters-suffice}
Let $\delta \leq \frac{1}{100}$ and $k \in \Z$ be large enough. Suppose the optimal SDP solution vectors $\set{u_i}_{i \in V}$ to \prettyref{sdp:primal-edge} (resp. \prettyref{sdp:primal}) yield an objective value of $\beta n$ and satisfy the following properties:
\begin{enumerate}
\item[(a)] There exist disjoint sets $L_1, L_2, \ldots, L_k \subseteq V$, with $\diam(L_t) \leq \delta$,
\item[(b)] For each $t \in [k]$, and for some constant $\gamma$, we have $\abs{L_t} \geq \gamma n/k$,
\item[(c)] For every $t \neq t'$, $d(L_t, L_{t'}) \geq 1/10$.
\end{enumerate}
Then, we can in polynomial time, find  $k$ disjoint sets $W_1, \ldots, W_k \subseteq V$  such that for every $t \in [k]$,  $\abs{W_t} \geq \gamma n/k$, and $\phi(W_t) \leq \bigO(\beta k/\gamma)$ (resp. $\phiv(W_t) \leq \bigO(\beta k /\gamma)$).
\end{lemma}

\begin{proof}
Let $\Phi = \phi$, if we are working with \kparte, and $\Phi = \phiv$ if we are working with \kpartv. The proof will work for either case. We first apply Lemma~\ref{lem:cluster-suffices} with $\alpha = \gamma/k$ to each of the sets $L_1, \ldots, L_k$ in turn to conclude the existence of the corresponding $W_1, \ldots, W_k$ sets each with vertex expansion at most $\bigO(\beta k/\gamma)$ .  Fix any $t \in [k]$. Note that from Lemma~\ref{lem:cluster-suffices}, the structure of $W_t$ implies that we have $L_t \subseteq W_t$, and hence $\abs{W_t} \geq \gamma n/k$ .  

\smallskip
Given the separation condition $(c)$, the sets $W_t$ are disjoint. Indeed, for any $t \neq t'$, if  $W_t = B(i,a)$, and  $W_{t'}=B(i',a)$ (where $i, i' ,a$ are given by \prettyref{lem:cluster-suffices}), we have, by the $\ell_2^2$ triangle inequality:
\begin{equation} \label{eq:W-separation}
d(W_t, W_{t'}) \geq d(i,i')-\frac{1}{50}-\frac{1}{50} \geq \frac{1}{20}
\end{equation}

\smallskip

Note that the above only shows \emph{existence} of $k$ disjoint sets. In order to actually find $k$ sets satisfying the given  conditions, we proceed greedily (this is the loop in step $2$ of Algorithm 1). At the first step, we find:
\[
\hat{W}_1 = \argmin  \set{ \Phi( B(i,a) )  \, : \, i \in V,  r\in [\delta, 1/50), \abs{B(i,a)} \geq \gamma n/k}
\]

Clearly, since $W_1$ is a candidate in the above minimization, we have $\Phi(\hat{W}_1) \leq \ \Phi(W_1) \leq \bigO(\beta k/\gamma)$. Furthermore, since $\diam(\hat{W_1})\leq 1/25$, \prettyref{eq:W-separation} implies that $\hat{W}_1$ can intersect at most one of the $W_t$ sets. This is because if $\hat{W_1}$ contained points from $W_t$ and $W_{t'}$, for $t \neq t'$, then \prettyref{eq:W-separation} implies that $\diam(W_1) \geq 1/20$, which is not possible. 

\smallskip

Now, we proceed similarly for $(k-1)$ more steps: at each step $t\in \set{2, \ldots k}$, find a set $\hat{W}_t$ that is disjoint from the previous $\hat{W}_1, \ldots, \hat{W}_{t-1}$ and has minimum $\Phi$. 

\[
\hat{W}_t = \argmin  \set{\Phi( B(i,r)) \, : \, i \in V,  r\in [\delta, 1/50), \abs{B(i,r)} \geq \gamma n/k, \hat{W_t}\cap \paren{\uplus_{s=1}^{t-1} \hat{W}_s} = \emptyset}
\]

\begin{lemma}\label{lem:feasible-Ws}
At the start of iteration $t$ in step 2 of Algorithm 1, there exists $A\subseteq [k]$, $|A| \geq (k-t+1)$ such that 
\[
\paren{\bigcup_{i \in A} W_t} \bigcap \paren{\bigcup_{i \in [t-1]} \hat{W_i}} = \emptyset  
\]
\end{lemma}
\begin{proof}
This is because, like $\hat{W}_1$, every subsequent $\hat{W}_t$ can intersect at most one of the  sets among $W_1, \ldots, W_k$. This implies that  at least $(k-t+1)$ of the $W_i$'s are \emph{untouched} at the start of iteration $t$, proving the statement of the lemma.
\end{proof}

From the above lemma, at every iteration in step 2 of Algorithm 1, there is always a $W_i$, for some $i \in [k]$ that is a feasible candidate for minimization at iteration $t$. This $W_i$ is known to satisfy the requirements on size ($\Omega(\gamma n/k)$) and expansion $\Phi (W_i) \leq \bigO(\beta k/\gamma)$. Thus, the above procedure always finds a non-empty $\hat{W}_t$, whose size is at least $\Omega(\gamma n/k)$, and which has $\Phi(\hat{W_t}) \leq \bigO(\beta k/\gamma)$. 
\end{proof}

\subsection{Showing that the SDP solution is clustered}

We next show that for any input instance from the class \kparte~or \kpartv~with appropriate parameters, every feasible set of SDP solution vectors are clustered. Using \prettyref{lem:k-clusters-suffice}, we can then immediately conclude the proof of \prettyref{thm:kparte} and \prettyref{thm:kpartv}.

Our main technical result is the following proposition.

\begin{proposition}\label{prop:sdp-is-clustered-2}
Let $\set{u_i}_{i \in V}$ be the optimal solution \prettyref{sdp:primal-edge} (resp. \prettyref{sdp:primal}) for an instance $G$ from \kparteparams~(resp. \kpartvparams) with  $\eps kr^3/\lambda \leq 1/800$. Then, there exist sets $L_1, \ldots, L_k \subseteq V$ such that:
\begin{enumerate}
\item[(a)] $\diam(L_t) \leq 1/100$,   
\item[(b)] $\forall t \in [k] : ~ \abs{L_t \cap S_t }\geq  n/2k$,
\item[(c)] $\forall t \neq t': \quad d(L_t, L_{t'}) \geq 1/10$.
\end{enumerate}
\end{proposition}

\begin{proof}[Proof of Proposition~\ref{prop:sdp-is-clustered-2}]

We begin by proving the following lemma.

\begin{lemma}\label{lem:sdp-is-clustered-1}
Let $\set{u_i}_{i \in V}$ be the optimal solution to the SDP for an instance $G$ from \kpartv~ or \kparte. For each $t\in [k]$, let $\mu_t = \E_{i \in S_t} [u_i]$. The following holds:

\begin{enumerate}
\item[(a)] $\forall t \in [k]:\quad \E_{j \in S_t} [ \norm{\mu_t - u_j}^2] \leq \frac{k\eps r^3}{\lambda}$
\item[(b)] $\quad 1\geq \norm{\mu_t}^2 \geq 1- k\epsilon r^3/\lambda$
\item[(c)] $\forall t \neq t' \quad \mu_t^T \mu_{t'} \leq k\eps r^3/\lambda$
\end{enumerate}
\end{lemma}
\begin{proof}
For this proof, we first discard/ignore all edges added by the monotone adversary within each $S_t$. We can do this without introducing errors, as an adversary adding edges within $S_t$ only increases $\sum_{ij\in E(S_t)} \norm{u_i - u_j}^2$. The proof only requires an upper bound on this quantity to work. This is argument is similar to that used in ~\cite{LV18} for handling a monotone adversary.

We require the following proposition regarding edge expander graphs; a proof can be found in \cite[Proposition 2.16]{LV18}.

\begin{proposition}[{{See \cite[Proposition 2.16]{LV18}}}]\label{prop:expansion-property}
Let $G'=(V',E')$ be an $n$-vertex edge-expander graph with spectral gap $\lambda$. Suppose that the degrees of the vertices in $G$ satisfy $\Delta(i) \in [d, rd]$, for some $r>1$, and $d \in \mathbb{N}$. Then for any $X \in \R^n$, we have:
\begin{equation}
\sum_{\set{i,j} \in E'} (X_i - X_j)^2 ~\geq~ \frac{1}{r^2} \cdot \frac{\lambda d}{n} \sum_{i,j \in V' \times V'} (X_i - X_j)^2  
\end{equation}
\end{proposition}

We will also need the following fact, the simple proof appears at the end of the current proof.
\begin{fact} \label{fct:centroid}
Let $\mu$ be the centroid of points $x_1, \ldots, x_N \in \R^n$. Then,
\[
\frac{1}{N^2} \sum_{i < j } \norm{x_i - x_j}^2 \,\, = \E_{i \in [N]} \norm{\mu - x_i}^2
\]
\end{fact}

\begin{proof}
Consider the points $y_i = x_i - \mu$, so $\sum_i y_i = \overrightarrow{0}$. Now, we have:
\[ 2 \cdot  \sum_{i<j} \norm{x_i - x_j}^2 =  \sum_{i,j} \norm{x_i - x_j}^2 = \sum_{i,j} \norm{y_i - y_j}^2 = 2 N \sum_{i} \norm{y_i}^2  = 2 N \sum_{i} \norm{\mu - x_i}^2  \] 
\end{proof}

\noindent \textbf{Item (a):}

 We need slightly different proofs for \kparte~ and \kpartv~models for this.

\textbf{[\kparte]:}
Since the SDP value is at most $2\cdot \eps rd \cdot n $, we have for every $t\in [k]$: 
\begin{align*}
\sum_{\set{i,j} \in E} \norm{u_i - u_j}^2 & \leq 2\cdot \eps rd \cdot n \\
\implies \sum_{\set{i,j} \in E(S_t)}\norm{u_i - u_j}^2 & \leq 2\cdot \eps rd \cdot n \\
\implies  \sum_{i,j \in S_t} \norm{u_i - u_j}^2 &\leq  2\cdot \eps n \cdot rd \cdot \frac{ n r^2}{k\lambda d} \\& \qquad\qquad \ldots \text{ using \prettyref{prop:expansion-property} within $S_t$, and $\abs{S_t} = n/k$}\\
\implies \E_{i \in S_t} \norm{\mu_t - u_i}^2 &\leq \frac{k \eps r^3}{\lambda}  \quad \ldots \text{ since $\mu_t$ is the centroid of $S_t$ and using \prettyref{fct:centroid}}
\end{align*}

\textbf{[\kpartv]:}
Since the SDP objective is $\sum_{i \in V} \eta_i \leq 2 \eps n$, we have:
\[
\forall t \in [k]: \quad \sum_{i \in S_t} \eta_i \leq 2 \eps n
\]

Fix some $t \in [k]$. Let  the degree of the edge expander within $S_t$ be in the range $[d, rd]$ for some integer $d$. Recall that $\eta_i = \max_{ j \in N(i)} \norm{u_i - u_j}^2$.
This implies:
\begin{align*}
\sum_{i \in S_t} \max_{j \in N(i)} \norm{u_i - u_j}^2 &\leq 2 \eps n \\
\implies \sum_{i \in S_t} \frac{1}{rd}\sum_{j \in N(i)\cap S'_t} \norm{u_i - u_j}^2  &\leq 2\eps n\\ &\qquad \ldots \text{ since average $\leq \max$, and the max degree is $rd$} \\
\implies \sum_{\set{i,j} \in E(S_t)} \norm{u_i - u_j}^2 &\leq  \eps n\cdot rd  \\
\implies \sum_{i, j \in S_t} \norm{u_i - u_j}^2 &\leq \eps n \cdot rd \cdot \frac{ n r^2}{k\lambda d}\\ & \qquad \ldots \text{ using \prettyref{prop:expansion-property} within $S_t$, and $\abs{S_t} = n/k$} \\
\implies \E_{i \in S_t} \norm{\mu_t - u_i}^2 &\leq \frac{k \eps r^3}{\lambda} \\ & \qquad \ldots \text{ since $\mu_t$ is the centroid of $S_t$ and using \prettyref{fct:centroid}} 
\end{align*}


\noindent \textbf{Item (b):} Since all the vectors $\set{u_i}_{i \in V}$ are unit vectors, and $\mu_t$ is an average of a subset of these, we have that $\norm{\mu_t}^2 \leq 1$. For the lower bound:
\begin{align*}
\frac{k\eps r^3}{\lambda} &\geq \E_{j \in S_t}[ \norm{\mu_t - u_j}^2]\\
 &= \norm{\mu_t}^2 + \E_{j \in S_t}\insquare{\norm{u_j}^2} - 2 \E_{j\in S_t}[\mu_t^T u_j] \qquad \ldots \text{ expanding out the terms} \\
 &\geq \norm{\mu_t}^2 +1  - 2 \cdot \norm{\mu_t}^2 \\& \qquad \qquad \ldots \text{ since all $u_j$'s are unit vectors, and using $ \E_{j \in S_t}[u_j] = \mu_t$} \\
 &= 1- \norm{\mu_t}^2   
\end{align*}
Rearranging yields the required lower bound.
 
 \medskip
\noindent \textbf{Item (c):} We know from the primal SDP constraint that for every $i \in V$,  $\sum_{j\in V} u_i^Tu_j = n/k$. Fix some $t_0 \in [k]$. 
\begin{align*}
\sum_{i \in S_{t_0}} \paren{\sum_{j \in V} u_i^T u_j} ~&\leq~ \paren{\frac{n}{k}}^2 \\
\implies \sum_{i \in S_{t_0}} \sum_{t \in [k]} u_i^T \mu_t ~&\leq~ \frac{n}{k} &\ldots \text{ since $\mu_t$ is the centroid of $S_t$}\\ 
\implies \frac{n}{k} \norm{\mu_{t_0}}^2 ~+~  \frac{n}{k}\sum_{t: t\neq t_0} \mu_{t_0}^T \mu_t  ~&\leq~ \frac{n}{k} & \ldots \text{ since } ~\frac{n}{k} \cdot \mu_{t_0} = \sum_{i \in S_{t_0}} u_i \\
 \implies \sum_{t: t\neq t_0} \mu_{t_0}^T \mu_t ~&\leq~ \frac{k\eps r^3}{\lambda} &\ldots \text{ using item (b) from this lemma}
\end{align*} 

Since all the inner products $u_i^T u_j$ are non-negative, each of the inner products in the last line are non-negative, and hence, all of them are upper bounded by $k\eps r^3/\lambda$, proving item (c) of \prettyref{lem:sdp-is-clustered-1}.

\end{proof}

The above concludes the proof of \prettyref{lem:sdp-is-clustered-1}. We use this to prove \prettyref{prop:sdp-is-clustered-2}.  For each $t \in [k]$, define $L_t \defeq B(\mu_t, 1/400)$. Clearly, $\diam(L_t)\leq 1/100$.
\smallskip

Since the parameters for either \kpart~model are assumed to satisfy  $\eps kr^3/\lambda \leq 1/800$, we have that for every $t\in [k]$, item (a) from \prettyref{lem:sdp-is-clustered-1}  implies that $ \E_{j \in S_t} [ \norm{\mu_t - u_j}^2] \leq k\eps r^3/\lambda \leq 1/800$. We can now use Markov's inequality:

\begin{align*} 
\Pr_{j \in S_t}\left[\norm{\mu_t - u_j}^2 > \frac{1}{400} \right] & = \frac{\abs{S_t \setminus (L_t \cap S_t)}}{\abs{S_t}}   &\ldots \text { since $L_t \defeq B(\mu_t, 1/400)$}\\
\implies \frac{\abs{L_t \cap S_t}}{\abs{S_t}} &= 1 - \Pr_{j \in S_t}\left[\norm{\mu_t - u_j}^2 > \frac{1}{400} \right]  \\
&\geq 1-\frac{\E_{j \in S_t} [ \norm{\mu_t - u_j}^2]}{ 1/400}  = \frac{1}{2}\\
\implies  \abs{L_t \cap S_t} &\geq \frac{n}{2k}
\end{align*}

To prove item (c) of the lemma, we first prove the following claim:

\begin{Claim}\label{claim:centers-are-far}
\[ \forall t\neq t' \qquad \norm{\mu_t - \mu_{t'}}^2 \geq \frac{9}{10} 
\] 
\end{Claim}
\begin{proof}
\begin{align*}
\norm{\mu_t - \mu_{t'}}^2 &= \norm{\mu_t}^2 + \norm{\mu_{t'}}^2 - 2 \mu_t^T \mu_{t'}\\
& \geq 1- \frac{k\eps r^3}{\lambda} +1 - \frac{k\eps r^3}{\lambda} - 2 \times  \frac{k\eps r^3}{\lambda}  &\ldots \text { using Lemma~\ref{lem:sdp-is-clustered-1}}\\
& \geq 1- \frac{4k\eps r^3}{\lambda} \geq \frac{19}{20} >\frac{9}{10} & \ldots \text{ since $\frac{k\eps r^3}{\lambda} \leq \frac{1}{800}$ \mper}
\end{align*}
\end{proof}

From the definition of the sets $\set{L_t}_{t\in [k]}$,  we will use  the (plain Euclidean) triangle inequality and the above claim. Let $t \neq t'$. We know that $d(L_t, L_{t'}) = d(i,i')$ for some $i \in L_t$ and $i' \in L_{t'}$. Using this:

\begin{align*}
d(L_t , L_{t'}) &= d(i, i') \\
&=\norm{u_i - u_{i'}}^2 \\
& \geq \inparen{\norm{\mu_t - \mu_{t'}} - \norm{\mu_t - u_i} - \norm{\mu_t - u_{i'}}}^2 \\& \qquad \qquad \qquad\ldots \textrm{ by triangle inequality on the point sequence $\mu_t \rightarrow i \rightarrow i' \rightarrow \mu_{t'}$}\\
&\geq \paren{ \norm{\mu_t - \mu_t'} -\frac{1}{20} - \frac{1}{20}}^2 \qquad \textrm{ \ldots since $d(\mu_{t'}, i') , \, d(\mu_t, i) \leq \sqrt{\frac{1}{400}} = \frac{1}{20}$}\\
& \geq  \paren{\frac{9}{10} - \frac{1}{10}}^2 > \frac{1}{10} \mper
\end{align*}

\end{proof}


\begin{algorithm} \label{alg:alg-1}
\caption{Algorithm for rounding SDP solutions for \kpartv~(\kparte) instances}
\begin{algorithmic}[1]
\INPUT $G=(V,E)$ from \kpartparams~and an optimal SDP solution  $\inbrace{u_i}_{i\in V}$ on $G$
\OUTPUT Disjoint sets $W_1, \ldots, W_k \subseteq V$ with $\abs{W_t} \geq  n/2k$
\State $C\gets \emptyset$
\For{$t \in 1, \ldots k$}
    \State $W_t \gets \emptyset$
	\For{$i \in V$}
	\For{$r \in [1/100, 1/50)$} \Comment{ Can be done in a discrete fashion}
	\State $\hat{W}  \gets  B(i,r)$ 
	\State \textbf{If}  {$\abs{\hat{W}} < n/2k$  \textbf{~or~}  $\hat{W}\cap C \neq \emptyset$ } \textbf{ continue}
	\State (For \kparte): \textbf{If} {$W_t = \emptyset$ \textbf{~or~}  $\phi(W_t) > \phi(\hat{W})$}  \textbf{then} $W_t \gets \hat{W}$ \hspace{3cm} \phantom{Buffertext} (For \kpartv): \textbf{If} {$W_t = \emptyset$ \textbf{~or~}  $\phiv(W_t) > \phiv(\hat{W})$}  \textbf{then} $W_t \gets \hat{W}$  
    \EndFor
	\EndFor
	\State $C \gets C \cup W_t$
\EndFor
\State  \Return $W_1, \ldots, W_t$
\end{algorithmic} 
\end{algorithm}

Using the above, we now infer the proof of \prettyref{thm:kparte} and \prettyref{thm:kpartv}.

\begin{proof}[Proof of Theorem~\ref{thm:kparte} and Theorem~\ref{thm:kpartv}]
Consider the optimal SDP solution vectors $\set{u_i}_{i\in V}$ for an instance $G$ from \kparteparams~(resp. \kpartvparams), with the parameters satisfying the given conditions, and having an objective value of $\beta n$. Note that $\beta \leq \opt$, as the SDP is a relaxation.  Using \prettyref{prop:sdp-is-clustered-2}, we infer the existence of sets $L_1, \ldots , L_k$ satisfying the conditions given.  The SDP solution thus satisfies all the conditions of \prettyref{lem:k-clusters-suffice}, with $\delta = \frac{1}{100}$ and $\gamma = 1/2$, and therefore, we can find in polynomial time, $k$ disjoint subsets $W_1, \ldots, W_k$: $\abs{W_t} \geq n/2k$, and $\phi(W_t)\leq \bigO(\beta k)$, for every $t\in [k]$ for \kparte, or correspondingly $\phiv(W_t)\leq \bigO(\beta k)$ for \kpartv. 
Algorithm 1 describes the steps in the algorithm explicitly.
\end{proof}

\begin{proof}[Proof of Corollary~\ref{cor:approx-recovery-1}]

The proof for both parts uses a technique to move from disjoint sets to partitions used before, for instance in \cite{lot14,lm14a}. Since these works do it for edge expansion already, we state the proof for \kpartv~first.

\textbf{For \kpartv :} 
We start with the sets $W_1, \ldots, W_k$ from \prettyref{thm:kpartv}.  From the definition of $\phiv$, we have:
\begin{align*}
|\partial W_t| = \abs{N(W_t)} + \abs{N(V\setminus W_t)} &\leq \bigO(1) \cdot \opt \cdot k \cdot \frac{\abs{W_t}\abs{V\setminus W_t}}{n} = \bigO(k \cdot \opt \abs{W_t}) & \forall t\in[k]
\end{align*}

Define the partition $\cP = \set{P_1, \ldots, P_k}$ as follows:  $P_i = W_i$ if $i\neq k$, and $P_k = V \setminus \uplus_{i\in [k-1]} W_i $. 
Clearly, we have:
\[
\abs{\partial P_k} \leq \abs{\bigcup_{t=1}^{k-1} \partial{W_t}} \leq \bigO(k \cdot \opt \sum_{t=1}^{k-1}\abs{W_t} )\leq \bigO(kn \cdot \opt)
\]

 Above, the last inequality follows since the $W_t$'s are all disjoint. Since $\abs{P_k} \geq \Omega(n/k)$, and $\abs{V\setminus P_k} \geq \Omega(n)$, we infer that $\phivk{\cP} \leq \phiv(P_k) \leq \bigo{ k^2 \cdot \opt}$. 

\smallskip

\textbf{For \kparte :} The proof is very similar to the preceding one for \kpartv, except we work with edges. Again, from the definition of $\phi$, we have, for the sets given by \prettyref{thm:kparte}:
\begin{align*}
|E(\partial W_t)| &\leq \bigO(1) \cdot \opt \cdot k \cdot \frac{\abs{W_t} \abs{V \setminus W_t}}{n} = \bigO(k \cdot \opt \abs{W_t})
\end{align*}

As before, we define $\cP = \set{P_1, \ldots, P_k}$ as follows:  $P_i = W_i$ if $i\neq k$, and $P_k = V \setminus \uplus_{i\in [k-1]} W_i $.  From the above bound on $\abs{E(\partial W_t)}$, we get that: $$\abs{E(\partial P_k)} = \bigO(k \cdot \opt \sum_{t=1}^{k-1} |W_t|) = \bigO(kn \cdot \opt )$$, giving that $\phi^k(\cP) = \bigO( k^2 \cdot \opt)$.
\end{proof}
%
%

%


\paragraph*{Acknowledgements}
AL was supported in part by SERB Award ECR/2017/003296 and a Pratiksha Trust Young Investigator Award.
{\small
\bibliographystyle{alpha}
\bibliography{k-way-bib}

\newcommand{\etalchar}[1]{$^{#1}$}
\begin{thebibliography}{ABKK17}

\bibitem[ABH16]{abh16}
Emmanuel Abbe, Afonso~S Bandeira, and Georgina Hall.
\newblock Exact recovery in the stochastic block model.
\newblock {\em IEEE Transactions on Information Theory}, 62(1):471--487, 2016.

\bibitem[ABKK17]{abkk15}
Naman Agarwal, Afonso~S Bandeira, Konstantinos Koiliaris, and Alexandra Kolla.
\newblock Multisection in the stochastic block model using semidefinite
  programming.
\newblock In {\em Compressed Sensing and its Applications}, pages 125--162.
  Springer, 2017.

\bibitem[Alo86]{a86}
Noga Alon.
\newblock Eigenvalues and expanders.
\newblock {\em Combinatorica}, 6(2):83--96, 1986.

\bibitem[AM85]{am85}
Noga Alon and Vitali~D Milman.
\newblock $\lambda_1$, isoperimetric inequalities for graphs, and
  superconcentrators.
\newblock {\em Journal of Combinatorial Theory, Series B}, 38(1):73--88, 1985.

\bibitem[ARV09]{arv09}
Sanjeev Arora, Satish Rao, and Umesh~V. Vazirani.
\newblock Expander flows, geometric embeddings and graph partitioning.
\newblock {\em Journal of the ACM}, 56(2), 2009.
\newblock (Preliminary version in {\em 36th STOC}, 2004).

\bibitem[AS15a]{as15a}
Emmanuel Abbe and Colin Sandon.
\newblock Community detection in general stochastic block models: Fundamental
  limits and efficient algorithms for recovery.
\newblock In {\em IEEE 56th Annual Symp. on Foundations of Computer Science
  (FOCS), 2015}, pages 670--688. IEEE, 2015.

\bibitem[AS15b]{as15b}
Emmanuel Abbe and Colin Sandon.
\newblock Recovering communities in the general stochastic block model without
  knowing the parameters.
\newblock In {\em Advances in neural information processing systems}, pages
  676--684, 2015.

\bibitem[BFK{\etalchar{+}}11]{bfk11}
Nikhil Bansal, Uriel Feige, Robert Krauthgamer, Konstantin Makarychev,
  Viswanath Nagarajan, Joseph Naor, and Roy Schwartz.
\newblock Min-max graph partitioning and small set expansion.
\newblock In {\em Foundations of Computer Science (FOCS), 2011 IEEE 52nd Annual
  Symposium on}, pages 17--26. IEEE, 2011.

\bibitem[BHT00]{bht00}
Sergey Bobkov, Christian Houdr{\'e}, and Prasad Tetali.
\newblock {$\lambda_\infty$, Vertex Isoperimetry and Concentration}.
\newblock {\em Combinatorica}, 20(2):153--172, 2000.

\bibitem[Bop87]{b87}
Ravi~B. Boppana.
\newblock Eigenvalues and graph bisection: An average-case analysis.
\newblock In {\em Proceedings of the 28th Annual Symposium on Foundations of
  Computer Science}, SFCS '87, pages 280--285, Washington, DC, USA, 1987. IEEE
  Computer Society.

\bibitem[BP99]{bp99}
Roberto Battiti and Marco Protasi.
\newblock Approximate algorithms and heuristics for max-sat.
\newblock In {\em Handbook of Combinatorial Optimization: Volume1--3}, pages
  77--148, Boston, MA, 1999. Springer US.

\bibitem[CLTZ18]{cltz18}
T.{-}H.~Hubert Chan, Anand Louis, Zhihao~Gavin Tang, and Chenzi Zhang.
\newblock {Spectral Properties of Hypergraph Laplacian and Approximation
  Algorithms}.
\newblock {\em J. {ACM}}, 65(3):15:1--15:48, 2018.

\bibitem[FHL08]{fhl08}
Uriel Feige, MohammadTaghi Hajiaghayi, and James~R. Lee.
\newblock Improved approximation algorithms for minimum weight vertex
  separators.
\newblock {\em SIAM Journal on Computing}, 38(2):629--657, 2008.

\bibitem[FK01]{fk01}
Uriel Feige and Joe Kilian.
\newblock Heuristics for semirandom graph problems.
\newblock {\em Journal of Computer and System Sciences}, 63(4):639--671, 2001.

\bibitem[GV16]{gv16}
Olivier Gu{\'e}don and Roman Vershynin.
\newblock Community detection in sparse networks via grothendieck’s
  inequality.
\newblock {\em Probability Theory and Related Fields}, 165(3-4):1025--1049,
  2016.

\bibitem[HLL83]{hll83}
Paul~W Holland, Kathryn~Blackmond Laskey, and Samuel Leinhardt.
\newblock Stochastic blockmodels: First steps.
\newblock {\em Social networks}, 5(2):109--137, 1983.

\bibitem[JS98]{js98}
Mark Jerrum and Gregory~B Sorkin.
\newblock The metropolis algorithm for graph bisection.
\newblock {\em Discrete Applied Mathematics}, 82(1):155--175, 1998.

\bibitem[KK95]{kk95}
George Karypis and Vipin Kumar.
\newblock Analysis of multilevel graph partitioning.
\newblock In {\em Proceedings of the 1995 ACM/IEEE Conference on
  Supercomputing}, Supercomputing '95, New York, NY, USA, 1995. ACM.

\bibitem[KK98]{kk98}
George Karypis and Vipin Kumar.
\newblock A fast and high quality multilevel scheme for partitioning irregular
  graphs.
\newblock {\em SIAM J. Sci. Comput.}, 20(1):359--392, December 1998.

\bibitem[KLL{\etalchar{+}}13]{kll13}
Tsz~Chiu Kwok, Lap~Chi Lau, Yin~Tat Lee, Shayan Oveis~Gharan, and Luca
  Trevisan.
\newblock Improved cheeger's inequality: Analysis of spectral partitioning
  algorithms through higher order spectral gap.
\newblock In {\em Proceedings of the Forty-fifth Annual ACM Symposium on Theory
  of Computing}, STOC '13, pages 11--20, New York, NY, USA, 2013. ACM.

\bibitem[LGT14]{lot14}
James~R Lee, Shayan~Oveis Gharan, and Luca Trevisan.
\newblock Multiway spectral partitioning and higher-order cheeger inequalities.
\newblock {\em Journal of the ACM (JACM)}, 61(6):37, 2014.

\bibitem[LM14]{lm14a}
Anand Louis and Konstantin Makarychev.
\newblock Approximation algorithm for sparsest k-partitioning.
\newblock In {\em Proceedings of the twenty-fifth annual ACM-SIAM symposium on
  Discrete algorithms}, pages 1244--1255. SIAM, 2014.

\bibitem[LM16]{lm16}
Anand Louis and Yury Makarychev.
\newblock Approximation algorithms for hypergraph small-set expansion and
  small-set vertex expansion.
\newblock {\em Theory of Computing}, 12(1):1--25, 2016.

\bibitem[LR99]{lr99}
Tom Leighton and Satish Rao.
\newblock Multicommodity max-flow min-cut theorems and their use in designing
  approximation algorithms.
\newblock {\em J. ACM}, 46(6):787--832, November 1999.

\bibitem[LRTV11]{lrtv11}
Anand Louis, Prasad Raghavendra, Prasad Tetali, and Santosh Vempala.
\newblock Algorithmic extensions of cheeger’s inequality to higher
  eigenvalues and partitions.
\newblock In {\em Approximation, Randomization, and Combinatorial Optimization.
  Algorithms and Techniques}, pages 315--326. Springer, 2011.

\bibitem[LRTV12]{lrtv12}
Anand Louis, Prasad Raghavendra, Prasad Tetali, and Santosh Vempala.
\newblock Many sparse cuts via higher eigenvalues.
\newblock In {\em Proceedings of the forty-fourth annual ACM symposium on
  Theory of computing}, pages 1131--1140. ACM, 2012.

\bibitem[LRV13]{lrv13}
Anand Louis, Prasad Raghavendra, and Santosh Vempala.
\newblock The complexity of approximating vertex expansion.
\newblock In {\em Proc. of the 54th Annual Symp. on Foundations of Computer
  Science}, FOCS '13, pages 360--369, Washington, DC, USA, 2013. IEEE Computer
  Society.

\bibitem[LV18]{LV18}
Anand Louis and Rakesh Venkat.
\newblock {Semi-random Graphs with Planted Sparse Vertex Cuts: Algorithms for
  Exact and Approximate Recovery}.
\newblock In {\em 45th International Colloquium on Automata, Languages, and
  Programming (ICALP)}, pages 101:1--101:15, 2018.
\newblock Full Version at: \url{https://arxiv.org/abs/1805.09747}.

\bibitem[Mas14]{m14}
Laurent Massouli{\'e}.
\newblock Community detection thresholds and the weak ramanujan property.
\newblock In {\em Proc. of the 46th Annual ACM Symp. on Theory of Computing},
  STOC '14, pages 694--703, New York, NY, USA, 2014. ACM.

\bibitem[McS01]{m01}
Frank~D. McSherry.
\newblock Spectral partitioning of random graphs.
\newblock In {\em Proc. of the 42nd IEEE Symp. on Foundations of Computer
  Science (FOCS)}, pages 529--537, Washington, DC, USA, 2001. IEEE Computer
  Society.

\bibitem[MMV12]{mmv12}
Konstantin Makarychev, Yury Makarychev, and Aravindan Vijayaraghavan.
\newblock Approximation algorithms for semi-random partitioning problems.
\newblock In {\em Proc. of the 44th Annual ACM Symp. on Theory of Computing},
  STOC '12, pages 367--384. ACM, 2012.

\bibitem[MMV14]{mmv14}
Konstantin Makarychev, Yury Makarychev, and Aravindan Vijayaraghavan.
\newblock Constant factor approximation for balanced cut in the pie model.
\newblock In {\em Proc. of the 46th Annual ACM Symp. on Theory of Computing},
  STOC '14, pages 41--49, New York, NY, USA, 2014. ACM.

\bibitem[MMV16]{mmv16}
Konstantin Makarychev, Yury Makarychev, and Aravindan Vijayaraghavan.
\newblock Learning communities in the presence of errors.
\newblock In {\em 29th Annual Conference on Learning Theory}, volume~49 of {\em
  Proceedings of Machine Learning Research}, pages 1258--1291, Columbia
  University, New York, New York, USA, 23--26 Jun 2016. PMLR.

\bibitem[MNS14]{mns14a}
Elchanan Mossel, Joe Neeman, and Allan Sly.
\newblock Belief propagation, robust reconstruction and optimal recovery of
  block models.
\newblock In {\em Conference on Learning Theory}, pages 356--370, 2014.

\bibitem[MNS15]{mns15a}
Elchanan Mossel, Joe Neeman, and Allan Sly.
\newblock Consistency thresholds for the planted bisection model.
\newblock In {\em Proc. of the 47th Annual ACM Symp. on Theory of Computing},
  STOC '15, pages 69--75, New York, NY, USA, 2015. ACM.

\bibitem[MNS17]{mns15b}
Elchanan Mossel, Joe Neeman, and Allan Sly.
\newblock A proof of the block model threshold conjecture.
\newblock {\em Combinatorica}, 2017.

\bibitem[MPW16]{mpw16}
Ankur Moitra, William Perry, and Alexander~S Wein.
\newblock How robust are reconstruction thresholds for community detection?
\newblock In {\em Proceedings of the forty-eighth annual ACM symposium on
  Theory of Computing}, pages 828--841. ACM, 2016.

\bibitem[NR01]{nr01}
J~Naor and Yuval Rabani.
\newblock Tree packing and approximating fc-cuts.
\newblock In {\em Proceedings of the twelfth annual ACM-SIAM symposium on
  Discrete algorithms}, volume 103, page~26. SIAM, 2001.

\bibitem[PSZ17]{psz17}
R.~Peng, H.~Sun, and L.~Zanetti.
\newblock Partitioning well-clustered graphs: Spectral clustering works!
\newblock {\em SIAM Journal on Computing}, 46(2):710--743, 2017.

\bibitem[RS08]{rs08}
R~Ravi and Amitabh Sinha.
\newblock Approximating k-cuts using network strength as a lagrangean
  relaxation.
\newblock {\em European Journal of Operational Research}, 186(1):77--90, 2008.

\bibitem[RS10]{rs10}
Prasad Raghavendra and David Steurer.
\newblock Graph expansion and the unique games conjecture.
\newblock In {\em Proceedings of the Forty-second ACM Symposium on Theory of
  Computing}, STOC '10, pages 755--764, New York, NY, USA, 2010. ACM.

\bibitem[RST10]{rst10}
Prasad Raghavendra, David Steurer, and Prasad Tetali.
\newblock Approximations for the isoperimetric and spectral profile of graphs
  and related parameters.
\newblock In {\em Proceedings of the Forty-second ACM Symposium on Theory of
  Computing}, STOC '10, pages 631--640, New York, NY, USA, 2010. ACM.

\bibitem[SV95]{sv95}
Huzur Saran and Vijay~V Vazirani.
\newblock Finding k cuts within twice the optimal.
\newblock {\em SIAM Journal on Computing}, 24(1):101--108, 1995.

\bibitem[WS11]{WS11}
David~P. Williamson and David~B. Shmoys.
\newblock {\em The Design of Approximation Algorithms}.
\newblock Cambridge University Press, New York, NY, USA, 1st edition, 2011.

\end{thebibliography}
}
\appendix

\section{Missing Proofs from Section~\ref{sec:approximate-recovery-proof}} \label{sec:appendix-approx-recovery}

\subsection{Proof of \prettyref{lem:cluster-suffices}}

Since the edge-expansion version is already well-known in literature, we start with vertex-expansion.
\begin{proof}[Proof of Lemma~\ref{lem:cluster-suffices}]

We start by stating the following lemma regarding $\ell_1$-line embeddings, a proof can be found in \cite[Appendix A.4]{LV18}. 

\begin{lemma}\label{lem:l1-embedding}
If there is a mapping $y:V \rightarrow \R$ that satisfies:
\[
\frac{ n \sum_{i} \max_{e=\{i,j\}}\abs{y_i - y_j}} {\sum_{i,j \in V} \abs{y_i - y_j}} \, = \, \delta_0  \mcom
\]
then there is a polynomial-time algorithm to find a non-trivial cut $(W, W')$ with vertex expansion at most $2\delta_0$. Furthermore, the set $W$ can be described using a threshold-cut: there exists some $t \in \R$ such that $W = \set{i: y_i \leq t}$.
\end{lemma}

The rest of the proof now closely follows the proof given for \cite[Lemma 3.1]{LV18}, requiring only minor modifications. We are given that there exists a set $L$ such that $\abs{L} \geq \alpha n$, satisfying:

\[
 \max_{i,j \in L } \norm{u_i - u_j}^2 \leq \frac{1}{100}
\]

Thus, we can fix an arbitrary $i_0 \in L$, which will satisfy :
\[
\abs{B\paren{i_0, \frac{1}{100}}} \geq \alpha n
\] 
 
Let $L' \defeq B\paren{i_0, \frac{1}{100}}$, so $\abs{L'} \geq \alpha n$. 

\begin{Claim}\label{claim:spread-constraint}
$\sum_{i,j \in V} \norm{u_i - u_j}^2 \geq 2 n^2(1-\frac{1}{k}) \mper$ 
\end{Claim}
\begin{proof}
\begin{align*}
 \sum_{i,j \in V} \norm{u_i - u_j}^2 & = \sum_{i,j} \paren{\norm{u_i}^2 + \norm{u_j}^2 - 2 \cdot u_i^T u_j} & \ldots \text{ expanding out the terms} \\
 & = \sum_{i,j} \paren{2 - 2 \cdot u_i^T u_j} &\ldots  \text { since $u_i$'s are unit vectors}\\
 & = 2n^2\paren{1- \frac{1}{k}} \mper  & \ldots \text{ since $\forall i$, } \sum_j u_i^Tu_j = \frac{n}{k} 
\end{align*}
\end{proof}

\begin{Claim}
For every $i \in V$, we have $\abs{B(i, 1/50)} \leq 9n/10$
\end{Claim}
\begin{proof}
Suppose $\abs{B(i, 1/50)} = m$. Since from the previous claim, $\sum_{i,j} d(i,j) = 2 \cdot n^2(1-1/k)$, we should have:
\begin{align*}
2n^2 \paren{1-\frac{1}{k}} &\leq m^2 \cdot \frac{1}{25} + (n^2 - m^2)\cdot 4 \\
\implies m^2 (4-\frac{1}{25}) &\leq \paren{2+\frac{2}{k}} n^2  \leq 3 n^2~\implies m \leq \frac{9}{10} \, n \mper
\end{align*}
The first line follows since pairs within the ball are at most $\frac{1}{25}$-squared distance apart due to the $\ell_2^2$ triangle inequality, while other pairs are at most squared distance $4$ apart, being unit vectors.
\end{proof}

Let $R' \eqdef V \setminus B(i_0, 1/50)$; from the above claim, we infer that $\abs{R'} \geq n/10$. Note that $L' \subseteq B(i_0, 1/100) \subseteq B(i_0, 1/50)$ and is hence disjoint from $R'$.  Furthermore, $d(L', R') \geq 1/50$ by the $\ell_2^2$ triangle inequality. Now, consider the mapping $y:V \rightarrow R^{+}$:
%

\[
y_i \eqdef 
\begin{cases}
\max \{0, d(i,i_0)- \frac{1}{100}\} & \text{for } i \notin R'\\
d(R',i_0) - \frac{1}{100}  & \text{for } i \in R' \mper
\end{cases}
\]

We show that the mapping $y$  satisfies the conditions of \prettyref{lem:l1-embedding}, with $\delta_0 = O(\beta)$.
\smallskip

We first show that $\abs{y_i - y_j} \leq d(i,j)$. To see this, consider three cases. First, say $i \notin R', j \in R'$. Clearly, $y_j = d(R', i_0) - 1/100 \leq d(j, i_0)-1/100$. Also, $y_i \leq y_j$, from the definition of $R'$. Hence, $\abs{y_i - y_j} = y_j - y_i \leq d(j, i_0) - d(i, i_0) \leq d(j,i)$, from the triangle inequality. 

Next, suppose that $i\notin R', j\notin R'$. Then $\abs{y_i - y_j} = \abs{d(i,i_0) - d(j, i_0)} \leq \abs{d(i,j)}$. The last inequality is from the $\ell_2^2$ triangle inequality.

Finally, suppose $i \in R', j \in R'$. Then $\abs{y_i - y_j} = {d(R', i_0)- d(R', i_0)} = 0 \leq d(i,j)$.  
%
%

Using the above, for any given $i$, we have:
\[
\max_{j\in N(i)}  \abs{y_i - y_j} \leq \max_{j\in N(i)} d(i,j)
\]
\smallskip

Next, we analyze the following sum:
\begin{align*}
\sum_{ij \in V} \abs{y_i - y_j} &\geq  \sum_{\substack{i \in L'\\ j \in R'}}  \abs{y_i - y_j} \\
& \geq \sum_{\substack{i \in L'\\ j \in R'}} \abs{0 - d(R', i_0) + \frac{1}{100}} \qquad \ldots \text{since $y_i = 0$, for $i \in L'$}\\  
& = \abs{L'} \abs{R'}\cdot \frac{1}{100}  \qquad \ldots \text{since $d(R', i_0) \geq 1/50$}\\
& \geq \Omega(\alpha) \cdot n^2 \\
& \geq \Omega(\alpha) \sum_{i,j} \norm{u_i - u_j}^2 \qquad \ldots \text{ from Claim~\ref{claim:spread-constraint}, using $k \geq 2$}
\end{align*}

Combining the above, we get that:
\[
\frac{ n \sum_{i} \max_{\{i,j\} \in E}\abs{y_i - y_j}} {\sum_{i,j \in V} \abs{y_i - y_j}} \, \leq  \, O(1) \cdot \frac{ n \sum_{i \in V} \max_{j \in N(i)} \norm{u_i - u_j}^2}{\alpha \sum_{ij} \norm{u_i - u_j}^2}
 ~\leq~ O(\beta/\alpha) 
\]

Using \prettyref{lem:l1-embedding}, we conclude that we can find a partition $(W, W')$  with vertex expansion $O(\beta / \alpha)$. From the nature of our embedding (a threshold cut with the threshold $\leq 1/50$), the required property on $W$ being $B(i_0, a)$ for some $a \in [\delta, 1/50]$ is clearly true. 


\medskip

\textbf{For \kparte :} The proof follows closely the above proof, so we only give a brief outline here. Instead of \prettyref{lem:l1-embedding}, we use a corresponding version for edge expansion that is well-known (see, for instance, the proof of \cite[Theorem 15.5]{WS11}).

\begin{lemma} \label{lem:l1-embedding-edge}
If there is a mapping $y:V \rightarrow \R$ that satisfies:
\[
\frac{ \sum_{i,j \in E} \abs{y_i - y_j}} {\sum_{i,j \in V} \abs{y_i - y_j}} \, = \, \delta_0 
\]
Then there is a polynomial-time algorithm to find a non-trivial cut $(W, W')$ with  $ \phi(W) \leq \delta_0$. Furthermore, the set $W$ can be described using a threshold-cut: there exists some $t \in \R$ such that $W = \set{i: y_i \leq t}$.
\end{lemma}

We now proceed exactly as in the vertex case, to get sets $L'$, $R'$ and the mapping $y_i$. Now, as we have that  $|y_i -y_j| \leq d(i,j)$, and $\sum_{ij} \abs{y_i - y_j } \geq \Omega(\alpha n^2)$, invoking \prettyref{lem:l1-embedding-edge} with $\delta_0 = \beta$, we recover the statement of the corollary.  
\end{proof}

\end{document}